\documentclass[aps,prb,reprint,showpacs,superscriptaddress]{revtex4-1}

\usepackage{amsmath}
\usepackage{amssymb}
\usepackage{bm}
\usepackage{setspace}
\usepackage{amsfonts}
\usepackage{rotating}
\usepackage{amsthm}
\usepackage[english]{babel}

\renewcommand{\vec}[1]{\bm{#1}}

\begin{document}

\title{Spin-orbit interaction in 2D Dirac-like and Kane semiconductors}
\author{E.~L.~Rumyantsev}
\author{P.~E.~Kunavin}
\affiliation{Institute of Natural Sciences, Ural Federal University, 620000 
Ekaterinburg, Russia}

\begin{abstract}
The single particle equations describing motion of carriers in external 
potential in 2D Dirac-like and Kane intrinsic semiconductors are obtained 
within second quantization method. The terms renormalizing external potential 
in these equations, referred to as spin-orbit (SO) terms, are compared with 
their classical counterpart. The well-known expression for SO obtained in 
relativistic Dirac theory arises in considered approach in the second order in
$\gamma k / E_g$ ($\gamma$ - characteristic velocity, $E_g$ - energy gap) 
parameter if electron-hole pair production terms are neglected. It is shown 
that in Kane problem the modifying terms are of standard SO functional 
Dirac-like form only for electrons in the case of ``positive'' energy gap and 
for light holes in semiconductors with ``negative'' energy gap. The general 
expression for renormalizing terms has in all cases non-local character. The 
arising of correction terms to single particle potentials which do not depend 
on band parameters is demonstrated for 2D gapless Dirac problem (graphene) and 
for Kane model. The origin of such  ``topological'' terms is attributed to the 
presence of degenerate bands in considered problems.
\end{abstract}


\maketitle

\section{Introduction}
The classical SO coupling enters into Schrodinger Hamiltonian describing 
electron (positron) motion in external potential from a non-relativistic 
approximation to the Dirac equation \cite{Rel_Quant_Theory}. The famous Pauli 
term $ \frac{\hbar}{4m^2c^2}\vec{\sigma}\left[\vec{\triangledown} 
V(r)\times \hat{\vec{p}}\right]$ describes the renormalization of external 
potential acting on electron due to the presence of positron band. In order to 
describe kinetics of carriers in semiconductors in terms of elementary 
excitations - electrons/holes the single particle Schrodinger-like equations
are to be extracted from multiband $\vec{k} \cdot \vec{p}$  Hamiltonians. 
Usually this derivation follows the text book pattern used in relativistic Dirac 
theory \cite{Rel_Quant_Theory}. In what follows we are not going to discuss
``pros and cons'' of such approach but want to point out very important   
inconsistency of proposed derivation emerging at the start. This derivation 
relies on the validity of the assumption that some components of Dirac 
four-component wave function are responsible of and can be used for the 
description of electron/positron movement. Here we come up against the 
problem: even in the free of external forces space the movement of 
electron/positron by the definition is determined by all four components of wave 
function which are all indispensable for the description. The most striking 
example of such situation is the problem of velocity operator, known as 
Zitterbewegung \cite{Schrodinger_ZB}. The admixture of ``non-proper'' components 
results in ``nonphysical'' values of free electron velocity. One possible way 
to circumvent this problem is by making Foldy-Wouthuysent (FW) transformation, 
which allows to break the Dirac equation into separate equations for positive 
and negative energies \cite{Foldy}. But this approach suffers from drawback 
that FW transformation is not unique \cite{Non_loc_FWT}. The same kind of 
difficulties are encountered while trying to obtain bound states in the field 
of supercharged  nuclei in Dirac theory \cite{Rel_Quant_Theory}. 

It is commonly accepted that Dirac equation is not a single-particle equation 
and must be considered within quantum field theoretical approach. ``The problem 
of charges in a fixed potential is usually treated by the method of second 
quantization of the electron field, using the ideas of the theory of 
holes.''\cite{Feynman} The Dirac theory without second quantization is known to 
lead to interpretational problems and it is claimed that ``single particle'' 
paradoxes such as e.g. Klein effect are resolved if quantum field line approach 
is used \cite{Krekora}. Thus it seems logical to try to analyze the problem of 
single-particle equations in the field formulation. The more so, as only within 
such consideration the quantum number ``charge'' can be unambiguously ascribed 
to the given state described by multi-component wave function. Such approach 
allows to avoid along way possible arising of spurious superposition of negative 
and positive energy states as according to the famous statement ``No one has 
ever succeeded in producing a state which is a superposition of states with 
different charges''.\cite{PCT}

It was long established \cite{Katsnelson20073, Zawadzki_ZB} that 
``quasi-relativistic'' problems infest any multiband semiconductor $\vec{k} 
\cdot \vec{p}$ Hamiltonians. The reason lies in the fact that bulk band 
structure of semiconductors is frequently described within the framework of 
$\vec{k} \cdot \vec{p}$ theory by matrix Hamiltonians which take into account at 
least two bands. Small wonder that if one proceeds along classical lines 
allowing for interference of positive and negative energy states such artefacts 
as effect of Zitterbewegung or Klein paradox emerge \cite{Zawadzki_ZB}. 
According to seminal paper by L. Keldysh \cite{Keldysh} the problem of 
supercharged  nuclei transforms in semiconductors into the problem of deep 
levels of impurity 
centers.

In order to use electron/hole language in envelope function approximation we are 
to ascribe as in Dirac theory the quantum number ``charge'' to the 
single-particle eigenfunctions of appropriate multiband $\vec{k} \cdot \vec{p}$  
Hamiltonian. Thus it seems natural to carry out the quest for these equations on 
the field of second quantization. It must be noted that description of 
semiconductor properties within quantum field theoretical approach rare if ever 
appears in the literature \cite{Gusynin}. The ``inconvenience'' of such line 
of approach for construction of single-particle equations lies in the fact that 
such procedure can be carried out only in momentum representation. Thus instead 
of solving differential equations in coordinate space we are to solve integral 
equations. It must be mentioned that while this is to some extent new concept 
for solving multi-component envelope function problem, it was claimed that 
transition to momentum space allows to avoid "spurious" solutions occurring in 
standard integration schemes \cite{Winkler_momentum_space}. The second obstacle 
is that second quantized version of any $\vec{k} \cdot \vec{p}$ Hamiltonian is 
multparticle even when inter-particle interaction is not accounted for due to 
possibility of pair production induced by external potential. It must be 
noted that under the term ``external potential'' we understood not only 
perturbation applied from without but impurity potentials existing within as 
well.

In what follows we start first of all with derivation of single-particle 
equations within proposed approach in ``semiconductor'' Dirac theory 
considering Dirac Hamiltonian as an example of the simplest two band 
$\vec{k} \cdot \vec{p}$ Hamiltonian. The obtained renormalized potential for 
Dirac-like semiconductors is compared with the classical (experimentally 
verified) expression for SO interaction, thus testing the validity of second 
quantization method for the solution of considered problem. Then we apply it to 
2D Kane semiconductors with ``positive'' and ``negative'' energy gap. 

It must be underlined that the term ``SO'' denoting sought correction to 
external potential in single-particle equations is used in the title and 
throughout the text in a broad sense. As it will be clear from the results 
presented below, physical origin and functional form of obtained corrections 
do not in general coincide with commonly used one.

\section{SO term in 2D Dirac-like semiconductor}
Leaving aside pure relativistic problems Dirac Hamiltonian is considered as an 
example of effective $\vec{k} \cdot \vec{p}$ Hamiltonian, describing two-band 
semiconductor with symmetrical conduction and valence bands. The Dirac 
Hamiltonian has the following form
\begin{equation}
\label{Dirac_H}
  \hat{H}_D =
   \begin{pmatrix}
    \frac{E_g}{2}  & 0     & \gamma k_z & \gamma k_-  \\
    0    & \frac{E_g}{2}   & \gamma k_+ & -\gamma k_z \\
    \gamma k_z & \gamma k_-  & -\frac{E_g}{2} & 0     \\
    \gamma k_+ & -\gamma k_z & 0    & -\frac{E_g}{2}
   \end{pmatrix} , 
\end{equation}
where  $k_{\pm} = k_x \pm ik_y$, $\gamma$ - characteristic velocity, $E_g$ - 
energy gap. In relativistic Dirac theory $E_g = 2mc^2$ and $\gamma = c$.  Here 
and in the following the atomic units ($\hbar = 1$) are used.

To simplify the following analysis, we consider 2D version of Hamiltonian 
\eqref{Dirac_H}. Choosing $k_z = 0$ two groups of 
states $(e ~1/2, h ~-1/2)$ and $(e ~-1/2, h ~1/2)$ do not mix. The Hamiltonian 
for the former group of these states reads
\begin{equation}
\label{Dirac_H_2D}
  \hat{H}_D = \begin{pmatrix}
              \frac{E_g}{2}  &  \gamma k_-   \\
              \gamma k_+    & -\frac{E_g}{2}
            \end{pmatrix} .
\end{equation}
The Hamiltonian matrix for the second group of states is obtained by replacing 
$k_y$ by $-k_y$. Therefore, we consider only first group of states described by 
Hamiltonian \eqref{Dirac_H_2D}.  The energy eigenvalues of 
\eqref{Dirac_H_2D} are
\begin{equation}
\varepsilon(k)_{1,2} = \pm\sqrt{\frac{E_g^2}{4} + \gamma^2 k^2} 
 \equiv \pm \varepsilon(k) .
\end{equation}

In the special case $E_g = 0$ Hamiltonian \eqref{Dirac_H_2D} is used for the 
description of electrons in $K$ valley in graphene
\begin{equation}
\label{Graphene_H}
  \hat{H}_K = \gamma
            \begin{pmatrix}
               0   & k_-   \\
               k_+ & 0
            \end{pmatrix} .
\end{equation}
But in this case we do not have two groups of spin states. Instead we are 
dealing with pseudospin, which is a formal way of taking into account the two 
carbon atoms per unit cell. The Hamiltonian for $K'$ valley is obtained from 
\eqref{Graphene_H} by making the transformation $H_{K'} = -H_K$.-

In the second quantization picture the Hamiltonian \eqref{Dirac_H_2D} 
for intrinsic semiconductor is
\begin{equation}
  \hat{H}_D = \int \varepsilon(k)\hat{a}^+(\vec{k})\hat{a}(\vec{k}) d\vec{k}
          + \int \varepsilon(k)\hat{b}^+(\vec{k})\hat{b}(\vec{k}) d\vec{k} \  .
\end{equation}
Here $\hat{a}^+(\vec{k}), ~\hat{a}(\vec{k})$ are creation/annihilation 
operators for electrons and $\hat{b}^+(\vec{k}), ~\hat{b}(\vec{k})$ are 
corresponding operators for holes. Inserting the potential $V(\vec{r})$ into
 “empty” Hamiltonian diagonal, in accord with the commonly accepted 
prescription, we obtain the following additional terms in the Hamiltonian
\begin{eqnarray}
\label{Dirac_H_int}
\nonumber
  \hat{H}_{i} &=&
   \int\int \varphi_e^*(\vec{k})\varphi_e(\vec{q}) V(\vec{k} - \vec{q})
     \hat{a}^+(\vec{k}) \hat{a}(\vec{q}){d\vec{k} d\vec{q}}\\
    \nonumber
    &-& 
    \int\int \varphi_h^*(\vec{k})\varphi_h(\vec{q}) V(\vec{k} - \vec{q})
     \hat{b}^+(\vec{q}) \hat{b}(\vec{k}){d\vec{k} d\vec{q}}\\
    \nonumber
     &+& 
    \int\int \varphi_e^*(\vec{k})\varphi_h(\vec{q}) V(\vec{k} - \vec{q})
     \hat{a}^+(\vec{k}) \hat{b}^+(\vec{q}){d\vec{k} d\vec{q}}\\
      &-& 
    \int\int \varphi_h^*(\vec{k})\varphi_e(\vec{q}) V(\vec{k} - \vec{q})
     \hat{a}(\vec{q}) \hat{b}(\vec{k}){d\vec{k} d\vec{q}} ,
\end{eqnarray}
where $\varphi_e(\vec{k})$, $\varphi_h(\vec{k})$ are eigen functions of 
\eqref{Dirac_H_2D} and
\begin{equation}
V(\vec{k}) = \int V(\vec{r})e^{-i\vec{k}\vec{r}}d\vec{r}.
\end{equation}
The terms containing $\hat{a}^+(\vec{k}) \hat{a}(\vec{q})$ 
and $\hat{b}^+(\vec{q}) \hat{b}(\vec{k})$ describe the processes of scattering 
electrons/holes by the potential modified by the presence of filled valence 
band. The terms containing $\hat{a}^+(\vec{k}) \hat{b}^+(\vec{q})$ and 
$\hat{a}(\vec{q}) \hat{b}(\vec{k})$ describe the perturbation of vacuum 
(ground) state which is the necessary attribute of second quantized 
consideration.

The explicit expression for modified electron scattering potential is
\begin{eqnarray}
\label{Dirac_Ve}
\nonumber
  V_e &=& \int\int N(k)N(q)\left( 1 + 
    \frac{4\gamma^2 k_- q_+}{(E_g + 2\varepsilon(k))(E_g + 2\varepsilon(q))}
    \right)\\
    \label{V_e}
     &\times& V(\vec{k} - \vec{q})  
      \hat{a}^+(\vec{k}) \hat{a}(\vec{q}){d\vec{k} d\vec{q}},
\end{eqnarray}
where $N(k) = \sqrt{\frac{E_g + 2\varepsilon(k)}{4\varepsilon(k)}}$.
The expression for holes (positrons) is of the same functional form but has 
opposite sign.

It can be easily shown, that expansion of \eqref{Dirac_Ve} up to the second 
order in $\gamma k/E_g \ll 1$ allows to present this expression as a sum of 
two parts
\begin{eqnarray}
\nonumber
  V_e &=& \int\int V(\vec{k}-\vec{q})\hat{a}^+(\vec{k}) 
\hat{a}(\vec{q}){d\vec{k} d\vec{q}}\\
   \nonumber
   &-&\frac{\gamma^2}{2E_g^2} \int\int \left( 
      2i(k_x q_y - k_y q_x) - (\vec{k} - \vec{q})^2 \right)\\
      \label{Dirac_Ve_initial}
      &\times&
       V(\vec{k}-\vec{q})
       \hat{a}^+(\vec{k}) \hat{a}(\vec{q}){d\vec{k} d\vec{q}}
\end{eqnarray}
The inequality $\gamma k/E_g \ll 1$ means that characteristic ``Compton'' 
wavelength $\lambda_C = \gamma/E_g$ is small as compared with the 
characteristic length of spatial variation of electron/hole envelope function 
$\lambda = 1/k$. The condition \mbox{$\lambda_C \ll \lambda$} is always implied 
while considering semiconductor problems within effective mass approximation. 
In coordinate representation expression \eqref{Dirac_Ve_initial} has the 
following form
\begin{equation}
 V_e(r) = V(r)
     + \frac{\gamma^2 }{E_g^2}
             \left[\vec{\triangledown} V(r)\times \hat{\vec{p}}\right]_z
     + \frac{\gamma^2}{2E_g^2}\vec{\triangledown}^2 V(r)
\end{equation}
It is seen that renormalized electron scattering potential contains well known 
SO interaction term as well as Darwin term. It must be stressed, that the 
description of electrons and holes dynamics separately using such renormalized 
potential is possible if we neglect the pair production terms induced by 
external potential.

In the special case $E_g = 0$ (graphene) modified external potential has 
rather simple form
\begin{eqnarray}
\nonumber
  \hat{H}_{i} &=&
   \frac{1}{2} \int\int \left(1+\frac{k_- q_+}{kq} \right) V(\vec{k} - \vec{q})
     \hat{a}^+(\vec{k}) \hat{a}(\vec{q}){d\vec{k} d\vec{q}}\\
    \nonumber
    &-& 
    \frac{1}{2} \int\int \left(1+\frac{k_+ q_-}{kq} \right) V(\vec{k} - \vec{q})
     \hat{b}^+(\vec{q}) \hat{b}(\vec{k}){d\vec{k} d\vec{q}}\\
    \nonumber
     &+& 
    \frac{1}{2}\int\int \left( \frac{k_+}{k} - \frac{q_-}{q} \right) 
     V(\vec{k} - \vec{q})
     \hat{a}^+(\vec{k}) \hat{b}^+(\vec{q}){d\vec{k} d\vec{q}}\\
      \label{Graphene_H_int}
      &-& 
    \frac{1}{2}\int\int \left(\frac{q_+}{q} - \frac{k_-}{k} \right) 
     V(\vec{k} - \vec{q})
     \hat{a}(\vec{q}) \hat{b}(\vec{k}){d\vec{k} d\vec{q}} .
\end{eqnarray}
The latter expression doesn't depend on any band parameters of semiconductor, 
but only on the problem's symmetry, thus demonstrating topological behavior 
\cite{Topolog_mater}.

In accord with the well-known result of Ref.~\onlinecite{Ando_backscattering}, 
it is seen from \eqref{Graphene_H_int} that back-scattering process ($\vec{k} = 
-\vec{q}$) is suppressed, as modified potential \mbox{$\left(1+\frac{k_\mp 
q_\pm}{kq} \right) V(\vec{k} - \vec{q})\Big|_{\vec{k} = -\vec{q}} \equiv 0$} for 
any potential, while pair production process is activated. This type of 
scattering process is similar to Andreev reflection \cite{Andreev}, in which 
initial electron passes the potential without scattering and simultaneously 
electron-hole pair is born.

Modified electron scattering potential \eqref{Dirac_Ve} clearly has a non-local 
character in coordinate representation. To investigate further its properties 
let us write it down for the linear potential of the form $V(\vec{r}) = F \cdot 
x$
\begin{eqnarray}
  \nonumber
  V_e &=& F \int\int \delta_{k_x, q_x}'(\vec{k}, \vec{q})
                   \hat{a}^+(\vec{k})\hat{a}(\vec{q}) d\vec{k}d\vec{q} \\
  \label{Dirac_Ve_linear}
          &-& F \int \frac{\gamma^2 k_y}{\varepsilon(k)(E_g + 2\varepsilon(k))} 
                   \hat{a}^+(\vec{k})\hat{a}(\vec{k}) d\vec{k} \ ,
\end{eqnarray}
where the linear potential in momentum representation is
\begin{eqnarray}
\nonumber
  V(\vec{k} - \vec{q}) 
   &=& F \cdot \frac{i}{2} \left\lbrace 
                    \delta_{k_x}'(\vec{k} - \vec{q}) 
                  - \delta_{q_x}'(\vec{q} - \vec{k}) 
                  \right\rbrace \\
   &=& F \cdot \delta_{k_x, q_x}'(\vec{k}, \vec{q}) \ .
\end{eqnarray}
The action of modifying term from \eqref{Dirac_Ve_linear} on a wave function of 
the form $\Psi(x,y) = exp(iq_y y)\varphi(x)$ in coordinate representation is as 
follows
\begin{equation}
  \hat{V}_{SO}\varphi(x) = \int K(x - x', q_y)\varphi(x')dx',
\end{equation}
where 
\begin{equation}
  K(\Delta x, q_y)
   = -F\int \frac{\gamma^2 q_y}{\varepsilon(k)(E_g + 2\varepsilon(k))}
           e^{ik_x\Delta x} dk_x \ .
\end{equation}
In the limit $ck/E_g \ll 1$ the asymptotical behavior of the kernel 
$K(\Delta x, q_y)$ is
\begin{equation}
  K(\Delta x, q_y) \sim \frac{\gamma q_y}{E_g}
   \left( \sqrt{2}e^{-\frac{E_g}{\sqrt{2}\gamma}|\Delta x|}
     - e^{-\frac{E_g}{\gamma}|\Delta x|} \right) ,
\end{equation}
thus the smearing is determined by ``Compton'' wave length $\gamma/E_g$. In the 
opposite limit $E_g \rightarrow 0$
\begin{equation}
  K(\Delta x, q_y) \sim e^{-|q_y||\Delta x|} ,
\end{equation}
i.e. the smearing is determined by de Broglie wave length $1/q_y$.

\section{SO term in 2D Kane semiconductor}
The proposed approach is applied for the analysis of narrow gap semiconductors, 
described by Kane Hamiltonian \cite{Kane1957249}. In order to make mathematics 
less complicated and more descriptive the simplified version of Kane Hamiltonian 
is used, accounting only for $\Gamma_6$ and $\Gamma_8$ bands, while neglecting 
remote bands contribution. In 2D version of this Hamiltonian for \mbox{$k_z = 
0$} two groups of states \mbox{$(e ~1/2, lh ~1/2, hh ~-3/2)$} and \mbox{$(e 
~1/2, lh ~-1/2, hh ~3/2)$} do not mix. The Hamiltonian matrix for the first 
group of these states in the case of "positive" energy gap is
\begin{equation}
\label{H_Kane_2D}
  \hat{H}_K = \begin{pmatrix}
                \frac{E_g}{2} & \frac{P}{\sqrt{6}}k_- & \frac{P}{\sqrt{2}}k_+ \\
                \frac{P}{\sqrt{6}}k_+ & -\frac{E_g}{2} & 0 \\
                \frac{P}{\sqrt{2}}k_- & 0 & -\frac{E_g}{2}
              \end{pmatrix} ,
\end{equation}
where $P$ - Kane's momentum matrix element. The Hamiltonian matrix for the 
second group of states is obtained by replacing $k_y$ by $-k_y$. The energy 
eigenvalues of \eqref{H_Kane_2D} are
\begin{gather}
  \varepsilon_{e,lh}(k) 
    = \pm\frac{1}{2}\sqrt{E_g^2 + \frac{8}{3}P^2 k^2} , \\
  \varepsilon_{hh}(k) = -\frac{E_g}{2} .
\end{gather}
In the considered approximation electrons and light holes (LH) have the same 
effective mass. This is similar to the Dirac Hamiltonian case, which also has 
symmetrical conduction and valence bands. But the main difference, which leads 
to striking change of SO terms, is the presence of heavy holes (HH) band, 
or more precisely the presence of degeneracy between LH and HH at $\Gamma$ 
point.

It must be stressed that we consider in some sense fanciful model, as there do 
not exist real 2D system described by Kane Hamiltonian. Nevertheless, it is 
reasonable to suppose that the main peculiarities inherent to real Kane systems 
are reproduced within such simplified model. The more realistic case of 
quasi 2D Kane systems (heterostuctures) will be considered elsewhere.

Using the same procedure as in Dirac semiconductor case, we obtain the 
following expression for modified electron scattering potential up to the 
second order in $Pk/E_g$ for ``positive'' gap situation
\begin{eqnarray}
\nonumber
  V_e &=& \int\int V(\vec{k}-\vec{q})\hat{a}^+(\vec{k}) 
\hat{a}(\vec{q}){d\vec{k} d\vec{q}}\\
   \nonumber
   &-&\frac{P^2}{3E_g^2} \int\int \left( 
      i(k_x q_y - k_y q_x) + (\vec{k} - \vec{q})^2 \right)\\
      \label{Kane_Ve_split}
      &\times&
       V(\vec{k}-\vec{q})
       \hat{a}^+(\vec{k}) \hat{a}(\vec{q}){d\vec{k} d\vec{q}}
\end{eqnarray}
It seen that the latter expression is similar to the one obtained for Dirac 
semiconductor. The coordinate representation equivalent of  
\eqref{Kane_Ve_split} is
\begin{equation}
 V_e(r) = V(r)
     + \frac{P^2 }{3E_g^2}
             \left[\vec{\triangledown} V(r)\times \hat{\vec{p}}\right]_z
     - \frac{P^2}{3E_g^2}\vec{\triangledown}^2 V(r)
\end{equation}
The obtained expression contains Rashba SO term \cite{Rashba}, as well as the 
term analogous to Darwin term. This result coincides with the one obtained in 
Ref.~\onlinecite{Winkler_SO_Coupl_eff}. Note one distinction of this expression 
from Dirac semiconductor. The prefactors ratio of SO and Darwin terms in the 
Dirac case is $2:1$, while in the Kane case it is $1:1$.

The situation with holes is more complicated. It has been already shown using 
purely group-theoretical methods that spin splitting for holes systems is very 
different from the spin splitting of electron states \cite{Winkler}. However we 
are not able to directly compare our results due to different problem's 
geometry chosen.

The crucial difference from Dirac semiconductor case within our approach comes 
from the fact that we have two types of hole states. This means that interaction 
part of Hamiltonian contains diagonal terms ($\hat{a}^+(\vec{k}) 
\hat{a}(\vec{q})$, $\hat{b}_{lh}^+(\vec{q})\hat{b}_{lh}$, 
$\hat{b}_{hh}^+(\vec{q})\hat{b}_{hh}$), pair production terms 
($\hat{a}^+(\vec{k})\hat{b}_{lh}^+(\vec{q})$, 
 $\hat{a}^+(\vec{k})\hat{b}_{hh}^+(\vec{q})$, 
$\hat{a}(\vec{k})\hat{b}_{lh}(\vec{q})$, 
$\hat{a}(\vec{k})\hat{b}_{hh}(\vec{q})$) as well as terms of the form 
$\hat{b}_{lh}^+(\vec{q})\hat{b}_{hh}(\vec{k})$ and 
$\hat{b}_{hh}^+(\vec{q})\hat{b}_{lh}(\vec{k})$. The latter terms describe 
the process of scattering between LH and HH states. Assuming the vacuum state 
to be stable, i.e. neglecting the electron-hole pair production terms, we 
obtain the following expression for the interaction part of effective 
Hamiltonian describing LH and HH behavior up to the second order in $Pk/E_g$
\begin{widetext}
\begin{eqnarray}
\nonumber
  H_i = &-& \int\int\left[\frac{2P^2}{3E_g^2}kq + \frac{1}{kq}
      \left( 1 - \frac{P^2}{3E_g^2}(k^2 + q^2) \right)
      \left( \vec{k}\cdot\vec{q} - \frac{i}{2}(k_xq_y - k_yq_x) \right) \right]
     V(\vec{k} - \vec{q})  
      \hat{b}_{lh}^+(\vec{q})\hat{b}_{lh}{d\vec{k} d\vec{q}} \\
\nonumber
     &-& \int\int \frac{1}{kq}
      \left( \vec{k}\cdot\vec{q} + \frac{i}{2}(k_xq_y - k_yq_x) \right)
     V(\vec{k} - \vec{q})
     \hat{b}_{hh}^+(\vec{q})\hat{b}_{hh}(\vec{k}){d\vec{k}d\vec{q}} \\
\nonumber
     &-& \int\int \frac{i\sqrt{3}}{2kq}\left( 1 - \frac{P^2k^2}{3E_g^2}\right)
     (k_xq_y - k_yq_x) V(\vec{k} - \vec{q})
     \hat{b}_{hh}^+(\vec{q})\hat{b}_{lh}(\vec{k}){d\vec{k}d\vec{q}} \\
\label{Kane_Hi}
     &-& \int\int \frac{i\sqrt{3}}{2kq}\left( 1 - \frac{P^2q^2}{3E_g^2}\right)
     (k_xq_y - k_yq_x) V(\vec{k} - \vec{q})
     \hat{b}_{lh}^+(\vec{q})\hat{b}_{hh}(\vec{k}){d\vec{k}d\vec{q}}.
\end{eqnarray}
\end{widetext}
It is seen that terms describing scattering between LH and HH states are of the 
same order as diagonal ones. As LH and HH states are of the same charge nothing 
forbids the interference between them which must be taken into account. Thus for
hole system the single-particle Hamiltonian does not exist and renormalized 
potential is described by $2 \times 2$ matrix (see \ref{Kane_Hi}). The 
importance of taking into account the mixing of these two types of holes has 
been underlined in Ref.~\onlinecite{Keldysh} while discussing the properties of 
deep levels in semiconductors.

Even the diagonal expression for modified LH $\rightarrow$ LH scattering 
potential qualitatively differs from the one obtained for electrons. It can't be 
easily split into usual Rashba SO and Darwin-like terms. To further investigate 
its properties let us write it down for the linear potential of the form 
$V(\vec{r}) = F \cdot x$
\begin{eqnarray}
  \nonumber
  V_{lh} &=& F \int\int \delta_{k_x, q_x}'(\vec{k}, \vec{q})
                   \hat{b}_{lh}^+(\vec{q})\hat{b}_{lh}(\vec{k})d\vec{k}d\vec{q} 
\\
  \label{Kane_Vlh_linear}
          &-& F \int \left[ -\frac{k_y}{2k^2} + \frac{P^2 k_y}{3E_g^2} \right] 
                   \hat{b}_{lh}^+(\vec{k})\hat{b}_{lh}(\vec{k}) d\vec{k} \ .
\end{eqnarray}
The latter expression contains usual SO term $\frac{P^2 k_y}{3E_g^2}$ as well 
as "topological" term $\frac{k_y}{2k^2}$, which is similar to the one 
discovered in Dirac semiconductor case for $E_g = 0$.

The diagonal expression for modified HH $\rightarrow$ HH scattering potential 
doesn't depend on band parameters of semiconductor and thus doesn't have usual 
expansion parameter $Pk/E_g$. In the case of linear potential it is
\begin{eqnarray}
  \nonumber
  V_{hh} &=& F \int\int \delta_{k_x, q_x}'(\vec{k}, \vec{q})
                   \hat{b}_{hh}^+(\vec{q})\hat{b}_{hh}(\vec{k}) d\vec{k}d\vec{q} 
\\
  \label{Kane_Vhh_linear}
          &-& F \int \frac{k_y}{2k^2} 
                   \hat{b}_{hh}^+(\vec{k})\hat{b}_{hh}(\vec{k}) d\vec{k} \ .
\end{eqnarray}
The latter expression contains the same "topological" $\frac{k_y}{2k^2}$ term 
as in LH case, but doesn't have usual SO term.

At this point we can conclude that the origin of "topological" terms is the 
presence of degeneracy between LH and HH states at $\Gamma$ point. This 
statement was also confirmed while considering Dirac-like semiconductor in the 
case $E_g = 0$, where the same type of terms were found. 

The Kane Hamiltonian for ``negative'' gap situation is obtained from 
\eqref{H_Kane_2D} by simple change of $E_g$ sign. Thus it is easy to show that 
in doing so the eigenfunctions for LH and electron states are to be formally 
permuted with $\vec{k} = -\vec{k}$, while functional form of HH eigenfunction 
remains the same. Although the functions index interchange is a trivial matter,
the physical consequences are far-reaching. As regards interaction of LH with 
external potential, it mimics the expression for electron SO in ``positive'' 
gap problem if electron-LH pair production is neglected
\begin{eqnarray}
\nonumber
  V_{lh} &=& -\int\int 
V(\vec{k}-\vec{q})\hat{b}_{lh}^+(\vec{q})\hat{b}_{lh}(\vec{k}){d\vec{k} 
d\vec{q}}\\
   \nonumber
   &+&\frac{P^2}{3E_g^2} \int\int \left( 
      i(k_x q_y - k_y q_x) + (\vec{k} - \vec{q})^2 \right)\\
      &\times&
       V(\vec{k}-\vec{q})
       \hat{b}_{lh}^+(\vec{q})\hat{b}_{lh}(\vec{k}){d\vec{k} d\vec{q}}.
\end{eqnarray}
As in the case of electron interaction with external potential in Kane 
semiconductor with ``positive'' gap the obtained expression is of the ``usual'' 
Rashba form. The renormalized potential acting on electron and HH states is
\begin{widetext}
\begin{eqnarray}
\nonumber
  H_i &=& \int\int\left[\frac{2P^2}{3E_g^2}kq + \frac{1}{kq}
      \left( 1 - \frac{P^2}{3E_g^2}(k^2 + q^2) \right)
      \left( \vec{k}\cdot\vec{q} - \frac{i}{2}(k_xq_y - k_yq_x) \right) \right]
     V(\vec{k} - \vec{q})  
      \hat{a}^+(\vec{k}) \hat{a}(\vec{q}){d\vec{k} d\vec{q}} \\
\nonumber
     &-& \int\int \frac{1}{kq}
      \left( \vec{k}\cdot\vec{q} + \frac{i}{2}(k_xq_y - k_yq_x) \right)
     V(\vec{k} - \vec{q})
     \hat{b}_{hh}^+(\vec{q})\hat{b}_{hh}(\vec{k}){d\vec{k}d\vec{q}} \\
\nonumber
     &-& \int\int \frac{i\sqrt{3}}{2kq}\left( 1 - \frac{P^2k^2}{3E_g^2}\right)
     (k_xq_y - k_yq_x) V(\vec{k} - \vec{q})
     \hat{a}^+(\vec{k}) \hat{b}_{hh}^+(\vec{q}){d\vec{k}d\vec{q}} \\
\label{Kane_Hi_neg}
     &-& \int\int \frac{i\sqrt{3}}{2kq}\left( 1 - \frac{P^2q^2}{3E_g^2}\right)
     (k_xq_y - k_yq_x) V(\vec{k} - \vec{q})
     \hat{a}(\vec{k}) \hat{b}_{hh}(\vec{q}){d\vec{k}d\vec{q}}.
\end{eqnarray}
\end{widetext}
It is seen that in this case the SO interaction for electrons can't be described 
by Rashba-like expression. Instead due to degeneracy of electron and HH 
bands scattering potential has ``topological'' terms contribution. Moreover pair 
production process can't be neglected, which makes this case similar to Dirac 
semiconductor case when \mbox{$E_g = 0$} (graphene).  Another difference from 
``positive'' gap case is that in this case scattering process between LH and HH 
states can be neglected since LH and HH band are separated by energy gap.

\section{Summary}
The derivation of Schrodinger-like single-particle equations describing 
electron/hole dynamics in external potential in intrinsic 2D Dirac and Kane 
semiconductors was carried out within second quantization method. Such approach 
allows automatically to take into account the effect of filled valence bands 
and to avoid in the cause of derivation unphysical superposition of positive 
and negative energy states. It is shown that obtained modified potential 
entering the sought equations may differ essentially from its classical ``SO'' 
counterpart as in its functional form, so in its dependence on band parameters. 
The degenerate bands modification of potential is very strong due to the 
presence of ``topological'' terms which do not depend on band parameters. Such 
strong modification of potential leads for instance to the absence of 
back-scattering in 2D Dirac problem, to the mixing of LH and HH states 
dependent only on the strength of potential in ``positive'' gap Kane 
problem and to the strong perturbation of valence vacuum in the proximity of 
e.g. impurity potential in the ``negative'' gap case. Moreover in coordinate 
representation the modified potential in Schrodinger-like equation, if it 
exists, is in general represented by non-local operator in all considered 
problems. This raises the interpretational issue. The generally accepted 
physical interpretation of SO origin is that it is a result of interaction of 
intrinsic electron/hole magnetic moment (spin) with magnetic field induced by 
orbital motion of the particle in external potential. Such ``classical'' 
interaction if of course presumed to be local. In our case the effective 
single-particle potential is renormalized in such a way as to exclude scattering 
of the electrons into occupied valence states. Thus it is reasonable to consider 
it as some kind of pseudo potential which actually does not differ in essence 
from the ones used in solid state physics. The presented approach show that SO 
terms understood as the renormalization of potential in  single-particle 
equations arises whenever we are dealing with two or more interacting energy 
bands. It is interesting that the similar statement was made in 
Ref.~\onlinecite{Zawadzki_ZB} while discussing the nature of Zitterbewegung. 
From this point of view the external potential renormalization is not necessary 
related to spin. E.g. consider graphene case where intrinsic SO interaction is 
extremely small \cite{Graphene_SO}, however the effect of external potential 
renormalization still takes place. As it was pointed above in order to remain 
within single-particle description (if possible) the terms leading to pair 
producing process must be disregarded. According to Schwinger \cite{Schwinger} 
the pair production process is activated when the work of external potential 
along Compton length exceeds $E_g$. For instance in a typical narrow gap 
semiconductor with $E_g = 100$ meV and $P = 8 \cdot 10^{-5} \text{meV} \cdot 
\text{cm}$ external electric field has to exceed $10^7$ V/m.

\section*{Acknowledgments}
Partial financial support from the RFBR (Grant No. 13-02-00322) is gratefully 
acknowledged.

\end{document}